\documentclass[aps,prb,preprint,superscriptaddress,showpacs]{revtex4}
\usepackage{graphicx}
\begin{document}
\title{Blue luminescence of SrTiO$_3$ under intense optical excitation}
\author{A. Rubano}
\affiliation{Dipartimento di Scienze Fisiche, Universit\`{a} di
Napoli ``Federico II'', Complesso di Monte S.Angelo, v.\ Cintia,
80126 Napoli, Italy}
\author{D. Paparo}
\author{F. Miletto Granozio}
\affiliation{CNR-INFM Coherentia, Complesso di Monte S.Angelo, v.\
Cintia, 80126 Napoli, Italy}
\author{U. Scotti di Uccio}
\author{L. Marrucci}
\email{lorenzo.marrucci@na.infn.it} \affiliation{Dipartimento di
Scienze Fisiche, Universit\`{a} di Napoli ``Federico II'', Complesso
di Monte S.Angelo, v.\ Cintia, 80126 Napoli, Italy}
\affiliation{CNR-INFM Coherentia, Complesso di Monte S.Angelo, v.\
Cintia, 80126 Napoli, Italy}
\begin{abstract}
The blue-green photoluminescence emitted by pure and electron-doped
strontium titanate under intense pulsed near-ultraviolet excitation
is studied experimentally, as a function of excitation intensity and
temperature. Both emission spectra and time-resolved decays of the
emission are measured and analyzed in the framework of simple
phenomenological models. We find an interesting blue-to-green
transition occurring for increasing temperatures in pure samples,
which is instead absent in doped materials. The luminescence yield
and decay rate measured as a function of temperature can be modeled
well as standard activated behaviors. The leading electron-hole
recombination process taking place in the initial decay is
established to be second-order, or bimolecular, in contrast to
recent reports favoring a third-order interpretation as an Auger
process. The temporal decay of the luminescence can be described
well by a model based on two interacting populations of excitations,
respectively identified with interacting defect-trapped (possibly
forming excitons) and mobile charges. Finally, from the measured
doping and sample dependence of the luminescence yield, we conclude
that the radiative centers responsible for the luminescence are
probably intrinsic structural defects other than bulk oxygen
vacancies.
\end{abstract}


\maketitle

\section{Introduction}
Strontium titanate, or SrTiO$_3$ (STO), is among the most widely
investigated perovskite oxides, owing both to its potential for
novel electronic applications and to its widespread use as a
substrate for the epitaxial growth of strongly-correlated electronic
materials, such as superconducting cuprates or
colossal-magnetoresistive manganites. Recently, the puzzling
transport properties of its interface with other insulating oxides
have also attracted much interest.\cite{huijben09,savoia09} Despite
this large effort, many properties of this material still await
complete clarification.

Intrinsic STO is a band insulator, characterized by a huge static
dielectric constant resulting from the rather soft bonding of the
small Ti$^{4+}$ ion to the surrounding octahedral O$^{2-}$ cage. Its
conduction band (CB) is composed of states having mainly Ti 3d
t$_{2g}$ character, while its valence band (VB) has dominantly O 2p
character, with an upper edge located away from the $\Gamma$ point
in the Brillouin zone.\cite{vanbenthem01} This results in an
indirect gap of 3.2-3.3 eV, while the direct (optical) gap is
3.4-3.7 eV, with a non negligible sample and temperature
dependence.\cite{vanbenthem01,capizzi70} Upon $n$-type doping,
generally achieved either by introducing O vacancies or by chemical
substitution (e.g., La$^{3+}$ for Sr$^{2+}$ or Nb$^{5+}$ for
Ti$^{4+}$), STO becomes a conductor with a relatively large
low-temperature mobility.\cite{frederikse67,keroack84} In this
regime, it is known that the charge carriers are dressed by the
interaction with the lattice and seem to behave mainly as large
polarons, although with several nonstandard
features.\cite{keroack84,eagles96,vanmechelen08,ishida08} When
intrinsic STO is irradiated with ultraviolet (UV) light,
photogenerated electrons and holes (e,h) appear to both contribute
to the material photoconductivity.\cite{itoh05} Theoretical
calculations would also indicate that holes in pure STO are not
strongly coupled to phonons and keep their bare
mass.\cite{stashans01}

The photoluminescence (PL) properties of STO are possibly even more
puzzling and controversial than its transport ones. The greenish
luminescence (GL) having a maximum at 2.2-2.4 eV of photon energy
(wavelength $\lambda\approx500$ nm) that is emitted by pure STO at
low temperature under exposure to UV or X-ray radiation has been
known since a long time,\cite{grabner69,feng82,aguilar82} and is
generally ascribed to the decay of intrinsic self-trapped excitons
(STE).\cite{leonelli86,hasegawa00,deguchi08,qiu08} A STE can be
roughly depicted as a tightly bound state of a hole and a Ti$^{3+}$
polaron.\cite{procel03,eglitis03} However, this purely intrinsic
scenario has been recently called into question by Mochizuki et
al.,\cite{mochizuki05,mochizuki06} who argued for a crucial role of
defects and possibly of surfaces in this GL emission.

Only recently, another photoluminescence emission taking place in
the blue (BL), with its maximum at 2.8-2.9 eV ($\lambda\approx430$
nm), was reported for STO at room
temperature.\cite{mochizuki05,kan05} This BL emission, potentially
useful for optoelectronic applications, is well visible both in
intrinsic samples at sufficiently high excitation
intensities\cite{mochizuki05} and in suitably $n$-doped samples, in
the latter case at much lower excitation
intensities.\cite{kan05,kan06} A similar blue luminescence was also
observed for intense electron-beam
excitation.\cite{grigorjeva04,zhang08} At low temperatures, the blue
emission is accompanied by a spectrally-narrow near-UV emission
(UVL) located at 3.2 eV ($\lambda\approx390$ nm), i.e.\ at band
edge.\cite{mochizuki05,kan05,kan06} Moreover, in some cases the BL
may also be accompanied by a long green ``tail'' covering a spectrum
similar to the GL discussed above, but still visible at room
temperature.\cite{mochizuki05,li07} It is not clear what determines
the appearance of this high temperature GL component, but surface
state of oxidation seems to play an important
role.\cite{mochizuki05} This GL is clearly visible at high
excitation intensities.\cite{rubano07,rubano08}

The underlying nature of this room temperature BL is still far from
clear. Although the BL yield is enhanced by $n$-doping, its spectrum
looks nearly identical for different kinds of doping as well as for
intrinsic samples under intense excitation (but this is only true at
room temperature, as we will see
below),\cite{mochizuki05,kan06,rubano08} thus pointing to a role of
conduction band electrons rather than donor levels in the
enhancement.\cite{kan06} When studied as a function of temperature,
in contrast with the GL which vanishes quickly above 30-50 K, the BL
exhibits a yield that is steadily increasing with temperature,
reaching a maximum at about 160 K and then decreasing slowly (a
faster decrease is however observed at temperatures substantially
higher than room temperature, as we will see
below).\cite{mochizuki05,yamada09} Studied as a function of the
excitation fluence (for short laser pulse excitation), the BL yield
shows no saturation up to very high fluences, in the mJ/cm$^2$
range, while the low-temperature GL has a much smaller saturation
threshold.\cite{mochizuki05,rubano07} More precisely, a detailed
recent study by Yasuda \textit{et al.} has shown that in pure STO
the BL yield is actually \textit{quadratic} in the excitation pulse
fluence up to about 1 mJ/cm$^2$.\cite{yasuda08} Above this value
there is initially a crossover to an approximately linear behavior
and then at 30-40 mJ/cm$^2$ a full
saturation.\cite{mochizuki05,rubano07,yasuda08} A strong dependence
of the BL spectrum and yield on a previous surface treatment with
fluorhydric acid has been also reported,\cite{li07} possibly
pointing to a role of the surface or of other crystal imperfections
in the luminescence process, although these results could be also
explained as resulting from a variation of the surface-induced
fluorescence quenching, thus unrelated with the radiative process
itself (in Ref.\ \onlinecite{li07} the excitation wavelength was 325
nm, at which the penetration length in STO is of only few tens of
nanometers, thus enhancing the role of the surface).

Even more intriguing is the PL dynamical behavior studied as a
function of time following a very short pulse excitation. It is
clearly established that the PL decay does not follow a simple
exponential behavior. At low temperatures, the GL band is associated
with a slow power-law decay, with a very strong temperature
dependence, typical of an untrapping-rate-limited ``bimolecular''
dynamics.\cite{leonelli86,hasegawa00,mochizuki05} The BL dynamics
has been studied in detail by us as a function of excitation energy
(in a strong excitation regime), and we observed an interesting
nonlinear dynamics, which can be modeled as a quite simple
two-component decay.\cite{rubano07} One component seems to be
exponential, or ``unimolecular'', while the other one can be fitted
well by a bimolecular power-law decay.\cite{rubano07} Although
apparently similar to the bimolecular behavior of GL seen at low
temperatures, the BL one is much faster (also at low temperatures)
and hence must be ascribed to a different process.\cite{mochizuki05}
The two dynamical components are present in the whole BL spectrum,
including its green ``tail'', apparently with no significant
wavelength dependence.\cite{rubano07} Doped samples exhibit a
similar two-component dynamics as pure ones and, at high excitation
fluences, also similar decay times and yields, with no evident
doping-induced enhancement.\cite{rubano08} However, as shown by
Yasuda \textit{et al.}, at smaller excitation intensities both the
overall luminescence yield and the exponential decay rate of the
unimolecular component are found to be strongly dependent on the
dopant concentration.\cite{yasuda08} In the same paper, in contrast
with our previous results, Yasuda \textit{et al.} claim that the
two-component decay is actually best described by an Auger
trimolecular process acting together with the unimolecular one, thus
leaving the issue of the leading recombination process governing the
PL decay undecided. Properly assessing the strength of the Auger
recombination may be important for proposed applications of STO in
opto-thermionic refrigeration.\cite{zhang09} We will come back to
this issue below.

In this work, we investigate further the physics underlying the blue
luminescence of STO by analyzing the PL spectra and temporal decays,
for both intrinsic and doped samples, as a function of excitation
energy and temperature. The temperatures of our measurements are
high enough to make the activated luminescence quenching evident.
Moreover, we compare in detail the predictions of different models
for the PL dynamical behavior, in order to establish the nature of
the recombination processes involved in the decay and to shed light
on the underlying electronic mechanisms.

This paper is structured as follows. Section 2 describes the
experimental procedures. The wavelength-resolved studies as a
function of excitation energy and temperature are reported and
analyzed, within a simple model, in Sec.\ 3. The corresponding
time-resolved studies are then reported in Sec.\ 4, together with a
first phenomenological modeling. In Sec.\ 5 we tackle the question
of a more detailed modeling of the dynamical behavior. The final
Section 6 includes a discussion of the physics underlying the
dynamical models used for interpreting the PL decay and a summary of
our main results.

\section{Experiments}
The samples used in this work were of the following three kinds: (i)
five stoichiometric intrinsic (100)-oriented STO single crystals
(I-STO), $1\times5\times5$ mm$^3$ in size, produced by four
different companies (SurfaceNet GmbH, CrysTec GmbH, Crystal GmbH,
eSCeTe B.V.) by the flame-fusion Verneuil method, with specified
impurity levels all below 150 ppm, and used as received; (ii) two
samples of Nb-doped STO crystals (N-STO), with a Nb molar
concentration of 0.2\%, and having the same orientation and geometry
as the I-STO samples; (iii) one sample doped with oxygen vacancies
(O-STO), obtained from a pure sample by annealing for 1 h at
950$^\circ$C and 10$^{-9}$ mbar (base pressure 10$^{-11}$ mbar).
While I-STO samples look transparent and are verified to be good
insulators, both N-STO and O-STO samples look black/dark-blue opaque
and are conducting. In O-STO, from resistivity measurements
performed on a thin film annealed by the same procedure, we estimate
an induced carrier density of $3\times10^{17}$ cm$^{-3}$, or about
0.002\%, rather small but still much higher than typical residual
oxygen vacancy concentrations in nominally stoichiometric
samples.\cite{rubano08} I-STO and N-STO samples did not show aging
or hysteretic behavior due to oxygen exchange with atmosphere, as we
checked by repeated measurements. Instead, O-STO samples turned
transparent again when heated above 250 $^\circ$C in air, clearly
showing re-oxidation of the vacancies. For this reason, all our
temperature behavior studies were limited to I-STO and N-STO
samples. During measurement, the samples were held into a thermostat
for temperature control to within 0.1 K (with an accuracy within few
K). The temperature was scanned between 300 K and 900 K.

In all our experiments, the excitation was induced by 3.49 eV UV
photons ($\lambda=355$ nm) in 25-ps-long laser pulses at 10 Hz
repetition rate focused to a gaussian spot having a radius of
$1.2\pm0.1$ mm at $1/e^2$ of maximum. The energy per pulse was
varied from 40 $\mu$J to 2 mJ, corresponding to an excitation
fluence $U$ ranging from a minimum of 2 to a maximum of 100
mJ/cm$^2$ (value at the spot center, corresponding to twice the
spatial-average value). This goes up to much higher values than
those investigated by others, including Mochizuki et al.\ (up to 14
mJ/cm$^2$)\cite{mochizuki05} and Yasuda \textit{et al.} (up to 10
mJ/cm$^2$)\cite{yasuda08}. For our highest fluence of 100 mJ/cm$^2$,
taking into account the 25\% reflection and assuming an optical
penetration length of about 1 ${\mu}m$,\cite{capizzi70} we estimate
a peak density of photogenerated e-h pairs as high as
$1.2\times10^{21}$ cm$^{-3}$. This corresponds to a
fluence-to-density conversion factor (FDCF)
$\alpha=1.2\times10^{22}$ cm$^{-1}$/J. It must be noted, however,
that the UV penetration length is highly uncertain, as different
samples have shown fairly different absorption edges in previous
reports. We stress that, despite our large excitation fluences, no
visible photoinduced damage of the sample surfaces was induced
during our experiments and we never observed irreversible variations
of the signal as a function of excitation intensity.

The luminescence emitted from the sample was collected by a lens
system imaging the illuminated sample spot onto the detector head,
after blocking the (much stronger) elastic scattering by a long-pass
filter with a cutoff wavelength of 375 nm. When recording the
luminescence spectra, for detection we used a grating-monochromator
and a photomultiplier and integrated the signal in time (with a 50
ns time gate). In time-resolved measurements, the luminescence was
instead detected with a photodiode (PD) having a rise-time of about
150 ps. In time-resolved experiments the entire luminescence
spectrum was integrated. The PD signal was acquired by a 20
Gsample/s digital oscilloscope having an analog bandwidth of 5 GHz.
The response function $r(t)$ of this apparatus was acquired by
measuring the signal given by the elastic scattering of the
excitation pulse (taken after removing the long-pass filter).

\section{Measurement results: spectra}
\begin{figure*}[t]
\includegraphics[angle=270, width=0.9\textwidth]{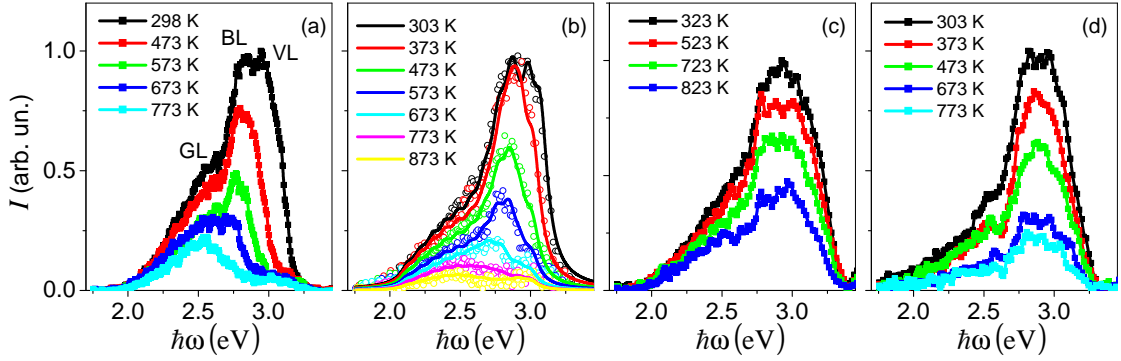}
\caption{(Color online) Photoluminescence spectra from I-STO (panels
a and b) and N-STO (panels c and d) samples at various temperatures
and at an excitation fluence of 2.2 mJ/cm$^2$ (panels a and c) or 22
mJ/cm$^2$ (panels b and d). Labels GL, BL, and VL (green, blue, and
violet luminescence) in panel (a) mark the peaks of the three
separate luminescence bands that can be singled out in our spectra.
A blue-to-green spectral transition (a red-shift of the peak) of the
photoluminescence is seen in the I-STO spectra, while it is absent
in the N-STO doped samples. The lines in panel (b) are best-fit
curves based on our theory. The lines in the other panels are guides
to the eye. \label{figspectra}}
\end{figure*}
The data shown in Fig.~\ref{figspectra} refer to a typical I-STO
(panels a and b) and a typical N-STO sample (panels c and d),
respectively under 2.2 (panels a and c) and 22 mJ/cm$^2$ (panels b
and d) excitation fluences, respectively. It is seen that the
overall shape of all spectra is asymmetrical, with both GL and BL
emissions present and clearly distinguishable at all temperatures,
in spite of a substantial overlap. At a closer inspection, an
additional spectral contribution apparently located at about 3.0 eV
($\lambda\approx400$ nm, in the violet), henceforth called VL (see
panel a), can be singled out, particularly in I-STO samples (it is
however possible that this emission is actually the same as the UVL
band mentioned in the Introduction, after a reshaping due to the
detection-line long-pass filter used in our setup). This VL appears
as a small shoulder in the lower temperature measurements, while it
is more clearly separated, although strongly depressed, at higher
temperatures. Our spectra are qualitatively consistent with
literature data. Three distinct emissions with similar spectral
position were reported in Ref.\ \onlinecite{mochizuki05}, though at
lower temperature. The spectrum of La-doped STO reported in Ref.\
\onlinecite{kan06} closely resembles that of our N-STO samples at
300 K. However, compared to previous data, ours seem to show a
comparatively higher yield in the GL region. In all samples the GL
contribution appears to increase more weakly than the BL one for an
increasing excitation intensity, as can be seen in
Fig.~\ref{figspectra} by comparing panels (a) with (b) and (c) with
(d). Therefore, the GL emission is probably more saturated than the
BL one, pointing to the presence of different classes of emitters as
responsible for the two bands.

As a function of temperature $T$, the most striking feature seen in
the PL spectra is a strong red-shift of the BL maximum occurring for
increasing $T$ in I-STO samples. I-STO samples at high temperature
emit almost exclusively in the green, with only a small residual
emission in the blue (which we ascribe to the VL band). This
blue-to-green thermal transition of the luminescence can be clearly
seen even with the naked eye. The fact that this shift does not
occur (or is very small) in N-STO doped samples, makes the spectra
of I-STO and N-STO samples become very different at high
temperatures. This is an important observation, as it is the first
clear qualitative difference seen in the luminescence spectra of
doped samples as compared to pure ones. In contrast to what stated
before,\cite{kan06,yasuda08} this shows that the role of doping is
not limited to introducing additional charge carriers in the system,
but somehow affects also the properties of the radiative and/or
non-radiative recombination centers.

\begin{figure}[t]
\includegraphics[angle=0, scale=0.9]{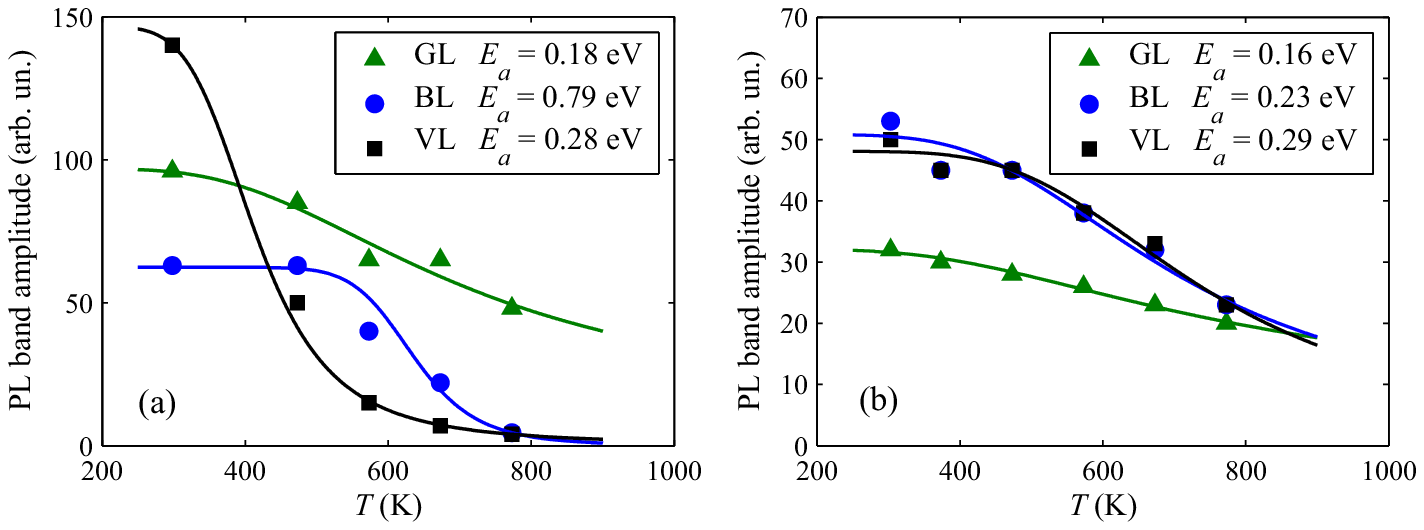}
\caption{(Color online) Temperature behavior of the amplitudes of
the three spectral bands GL (triangles), BL (circles), and VL
(squares) as seen in the I-STO (panel a) and N-STO (panel b)
photoluminescence. The excitation fluence was 2.2 mJ/cm$^2$. Solid
lines are best fits based on Eq.\ (\ref{eqact}); the resulting
activation energies are reported in the legend. The case of high
excitation fluence (22 mJ/cm$^2$) gives similar results, except for
a slightly smaller activation energy of the VL component
($\approx0.2$ eV instead of $\approx0.3$ eV).
\label{figbandamplitudes}}
\end{figure}

In order to analyze the data quantitatively, we will refer to a
specific model for the spectral shape $I(\omega)$ of each component
of the luminescence spectrum (GL, BL, and VL), treating them as
independent. As previously stated, the low-temperature GL was
interpreted in a quantitative way by Leonelli and Brebner in terms
of the annihilation of self-trapped excitons of given energy
$E_0$.\cite{leonelli86} In this picture, the broadening of the band
is due to the random emission of several optical phonons, each of
energy $\hbar\Omega\approx90$ meV, giving rise to the following
expression:
\begin{equation}
I(\omega)=I_0e^{-S_0}\sum_{n=0}^{\infty}\frac{S_0^n\Gamma/(2\pi n!)}
{(E_0-\hbar\omega-n\hbar\Omega)^2+(\Gamma/2)^2}, \label{eqspectra}
\end{equation}
whose key parameter is the Huang-Rhys factor $S_0$, related to the
strength of the electron-lattice interaction and fixing the average
number of emitted phonons per recombination. $\Gamma$ is the width
of the band associated with a given $n$-phonon process, taken to be
Lorentzian. Although devised for self-trapped excitons, Eq.\
(\ref{eqspectra}) applies equally well to the case of
defect-assisted recombinations -- in which defects mediate the
coupling of the electronic excitation to the lattice -- , so it may
be actually considered as a semi-phenomenological model that can
describe different microscopic scenarios. Leonelli and Brebner found
that a value as high as $S_0=5.7$ produced a nice fit to the data,
meaning that the maximum of intensity at 2.4 eV is well shifted with
respect to the intrinsic exciton energy, taken at $E_0\approx2.9$
eV. We plotted the $I(\omega)$ resulting from Eq.\
(\ref{eqspectra}), keeping the same quoted values of the parameters
$S_0$ and $E_0$ and assuming a broadening factor $\Gamma=0.12$ eV:
even with no adjustable parameters (except for the overall amplitude
scale), we could describe in this way quite accurately the GL band
shape in the region where it is well separated from the other bands
(see, e.g., the highest temperature spectrum in
Fig.~\ref{figspectra}b). Hence, we assumed that the GL band keeps
the same form in all spectra, except for a temperature dependent
scale factor, and turned to the problem of the BL and VL bands.
Tentatively, given its general applicability, we used Eq.\
(\ref{eqspectra}) also for these bands, and after suitable adjusting
of the parameters $E_0$ and $S_0$ we could achieve a satisfactory
fit of all data. The phonon energy $\hbar\Omega$ and the irrelevant
broadening factor $\Gamma$ were always kept fixed at the values 88
meV and 0.12 meV, respectively, without any attempt of optimization.
The optimal values of the characteristic energies and of the
Huang-Rhys factors for the GL, BL and VL components were determined
by a global fit procedure as $\simeq$2.9 eV, $\simeq$2.9 eV,
$\simeq$3.0 eV, and $\simeq$5.5, $\simeq$1, $\simeq$1, respectively,
independently of fluence and temperature, and allowing for only a
slight sample dependence. In addition, we used as free fit
parameters the respective amplitudes $I_0$ of the three bands at
each temperature. Typical results for pure and doped samples are
reported in Fig.~\ref{figbandamplitudes}. As a last step, we
observed that allowing for a slight decrease of the characteristic
energy of the VL contribution for increasing temperature (in any
case below 0.1 eV) the fit was further improved. This shift can be
probably associated with the known temperature dependence of the STO
gap.\cite{blazey71} An example of the fit results is given in
Fig.~\ref{figspectra}b. The overall fit quality is as good in all
I-STO and N-STO investigated samples.

Figure \ref{figbandamplitudes}a shows that the blue-to-green
transition of I-STO samples can be explained as a thermal quenching
of the VL and BL components occurring at lower temperatures than for
the GL one. This is not the case of N-STO samples, for which the
three amplitudes have a more similar behavior with temperature, as
shown in Fig.~\ref{figbandamplitudes}b. The thermal quenching of the
PL amplitudes $I_0(T)$ can be approximately modeled by the following
standard Arrhenius activation law:
\begin{equation}
I_0(T) \propto \tau(T) \propto \left[a+be^{-E_a/(k_bT)}\right]^{-1},
\label{eqact}
\end{equation}
where $\tau$ is the characteristic PL decay time, $E_a$ an
activation energy giving the potential barrier of the competing
non-radiative relaxation channels, $k_b$ the Boltzmann constant and
$a,b$ are constants that give the relative weight of the temperature
independent (typically radiative) and thermally activated (typically
non-radiative) contributions to the decay, respectively. The best
fit results based on Eq.\ (\ref{eqact}) are also shown in
Fig.~\ref{figbandamplitudes}. The corresponding activation energies
are all of the order of tenths of eV, with somewhat smaller values
for GL (0.16-0.18 eV), and slightly larger for BL and VL (0.23-0.29
eV), with the exception of the BL in I-STO, which is found to be
about 0.8 eV. According to these best-fit results, the stronger
thermal quenching of the BL and VL bands compared to GL is not
associated to a smaller activation energy (it is actually larger),
but to a relatively larger weight of the non-radiative channels with
respect to the radiative ones.

\begin{figure}[t]
\includegraphics[angle=0, scale=1]{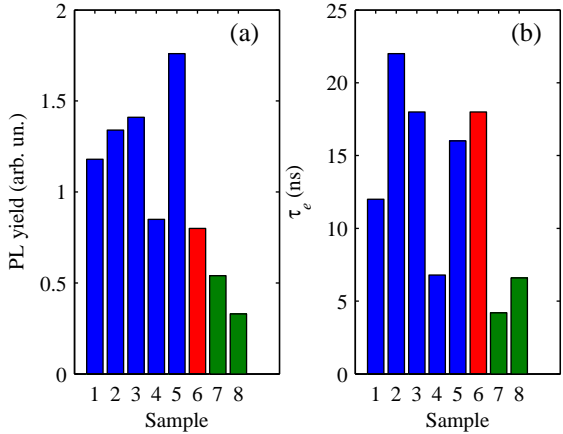}
\caption{(Color online) Histogram of the PL yield (panel a) and
PL-tail exponential decay time $\tau_e$ (panel b), for different
samples at room temperature. Samples 1-5 are I-STO (blue online),
sample 6 is O-STO (red online) and samples 7-8 are N-STO (green
online). \label{fighisto}}
\end{figure}
Finally, the spectrally-integrated PL yield at a pump fluence of 2.2
mJ/cm$^2$ as a function of the sample at room temperature is shown
in Fig.~\ref{fighisto}a. The most important things to notice here
are the following: (i) there is a significant sample-to-sample
dependence of the yield in nominally identical I-STO samples; (ii)
doped O-STO and N-STO samples do not show any enhanced yield,
despite the presence of many additional donor defects. We will
discuss these findings in Sec.\ 6.

\section{Measurements results: temporal decays}
\begin{figure}[t]
\includegraphics[angle=0, scale=0.95]{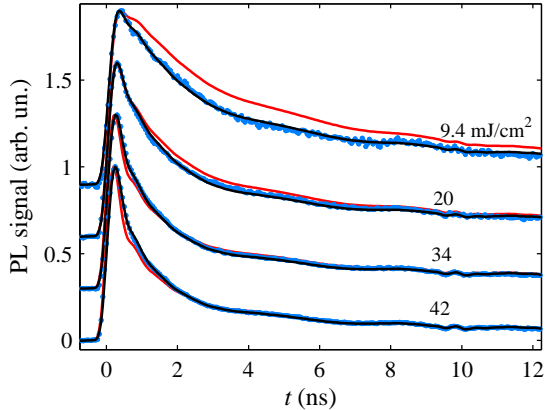}
\caption{(Color online) Time decay of the PL signal of a I-STO
sample following its excitation by a UV picosecond pulse, for
different excitation fluences, at room temperature. Data are shown
as gray dots (blue online). The solid lines are the result of a
global best fit based on our models (including a final convolution
with the measured instrumental response time $r(t)$). The black line
is based on the C2PUBv1 bimolecular model and the gray line (red
online) is based on the 1PUT trimolecular model. Data and curves
referring to different fluence values are vertically shifted for
clarity. \label{figconfrontodecays14STO}}
\end{figure}

\begin{figure}[t]
\includegraphics[angle=0, scale=0.95]{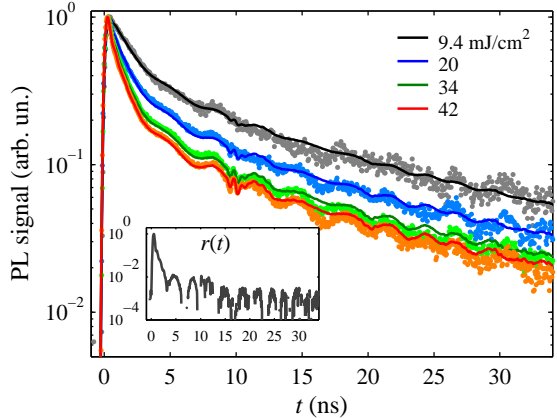}
\caption{(Color online) Time decays of the PL signal of a I-STO
sample at room temperature, for different excitation fluences, shown
here in semi-logarithmic scale to highlight the exponential PL tail.
Data are shown as dots, while the solid lines are the result of a
global best fit based on the C2PUBv1 model. The wiggles seen in the
tail for both data and model predictions are due to the
detection-system response function $r(t)$, which is shown in the
inset (inset axes labels and units are the same as for main
panel).\label{figdecaysech30log}}
\end{figure}

A typical set of PL temporal decays measured from a I-STO sample for
various excitation pulse fluences $U$ is shown in
Figs.~\ref{figconfrontodecays14STO}-\ref{figdecaysech30log}. The
case of N-STO is very similar (an example is reported in Fig.~2 of
Ref.\ \onlinecite{rubano08}).
As already noted in our previous works and confirmed by Yasuda
\textit{et al.},\cite{rubano07,rubano08,yasuda08} the PL initial
decay becomes faster for higher excitation energies, while the final
part of the decay (the ``tail'') varies only in its amplitude,
relative to the peak, but not in its rate. One can therefore
phenomenologically single-out two distinct regimes: the initial fast
decay, with an excitation-dependent characteristic decay time, and
the final tail, with a well defined, excitation-independent,
exponential decay time $\tau_e$. In I-STO, the fast decay is
typically in the range 1-2 ns (for our excitation fluences) and is
only weakly sample dependent, while the slow tail time constant
ranges from 6 to 23 ns, depending on the sample, as shown in
Fig.~\ref{fighisto}b. In N-STO the fast decay rate is of the same
order as in I-STO, while the exponential decay becomes somewhat
faster ($\tau_e\approx5$ ns), as an effect of doping.\cite{yasuda08}

In Figs.~\ref{figconfrontodecays14STO}-\ref{figdecaysech30log} the
decay signals are normalized to their maximum, for clarity. However,
also the PL signal amplitude varies strongly with the excitation
intensity. More precisely, as mentioned in the Introduction, we
observed an approximately linear dependence of the time-integrated
PL signal, i.e.\ of the overall PL yield, on excitation fluence up
to 30-40 mJ/cm$^2$, while for even higher fluences we find a
saturation (see Fig.~2 of Ref.\ \onlinecite{rubano07} and Fig.~3 of
Ref.\ \onlinecite{rubano08}). The PL signal maximum typically shows
a mixed linear-quadratic behavior with excitation fluence, with a
relatively large scatter of the data, for reasons which we have not
yet identified. The exact threshold for saturation is also sample
dependent, pointing again to an important role of the defects in the
PL radiative channels. Moreover, the linear yield behavior we find
in our data is actually already a partly saturated one, as for lower
excitation fluences a quadratic behavior is observed
instead.\cite{yasuda08}

\begin{figure}[t]
\includegraphics[angle=0, scale=0.95]{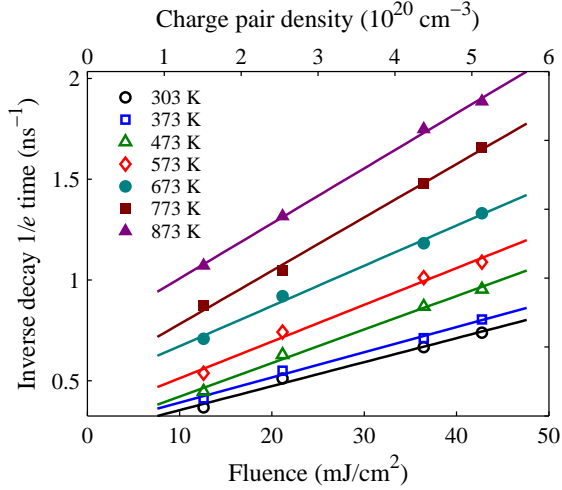}
\caption{(Color online) Experimental decay rates as a function of
excitation fluence and temperature, for a I-STO sample. The decay
rate is here defined as the inverse of the FW$1/e$M decay time
$t_{1/e}$. The upper $x$ axis gives our estimate of the density of
photoinduced electron-hole pairs after excitation. The lines are
linear best fits of the fluence dependence.\label{figrates}}
\end{figure}

To specify a characteristic experimental decay time for each given
PL signal, we use its full-width-at-$1/e$-of-the-maximum (FW$1/e$M)
time $t_{1/e}$. Equivalently, its inverse $t_{1/e}^{-1}$ provides a
characteristic decay rate. The measured decay rates $t_{1/e}^{-1}$
versus excitation fluence $U$, for different temperatures, in a
I-STO sample, are shown in Fig.~\ref{figrates}. N-STO samples give
very similar results.
In the investigated range, this decay rate shows an approximate
linear dependence on the excitation fluence, as shown in the figure.
However, since the measured decay times $t_{1/e}$ are close to the
characteristic response time of our detection apparatus, these raw
data for $t_{1/e}$ are significantly larger than the actual decay
times of the luminescence. To take care of this, in principle we
should deconvolve the measured decay signal and the response
function of our setup. However, the numerical deconvolution of noisy
data is known to be problematic, so that, following the standard
procedure, we used the inverse approach. Given a theoretical model
$I(t)$ for the PL decay containing the characteristic times to be
determined as adjustable parameters, we first convolve it with the
measured response function $r(t)$, thus obtaining a predicted signal
$S(t)=(r*I)(t)$. The latter is then compared with the measured
signal, thus finding the best-fit values of the adjustable
parameters. In this way, the best-fit values of the characteristic
decay times appearing as parameters in the model function $I(t)$
will correspond to actual PL decay times, without significant
distortions due to the instrumental response function. The final
time-resolution that can be achieved with this approach is limited
only by the signal-to-noise ratio of our data (typically
$10^2-10^3$). In this Section, we wish to analyze the data without
relying on too specific physical assumptions: ideally, we would like
to adopt a model-independent description, postponing the analysis of
specific models to the following Section. Therefore, we adopt here
the ``phenomenological'' model already used in Ref.\
\onlinecite{rubano07} (corresponding to the 2PUB considered in next
Section). According to this model, the PL intensity as a function of
time is given by the sum of a pure unimolecular exponential decay
with time constant $\tau_1$ and a pure bimolecular decay with
excitation-dependent time constant $\tau_2=1/(\gamma\alpha_2U)$,
where $\gamma$ and $\alpha_2$ are constants.\cite{rubano07} The fast
FW$1/e$M time $t_{1/e}$ will be approximately proportional to the
value of $\tau_2$. For each given sample and temperature, we may
assign a typical value to the time constant $\tau_2$ by fixing a
reference excitation fluence $U_1$ at which it must be evaluated. We
choose here a value for $U_1$ that is in the order of our
experimental range, i.e.\ $U_1=10$ mJ/cm$^2$.

\begin{figure}[t]
\includegraphics[angle=0, scale=0.9]{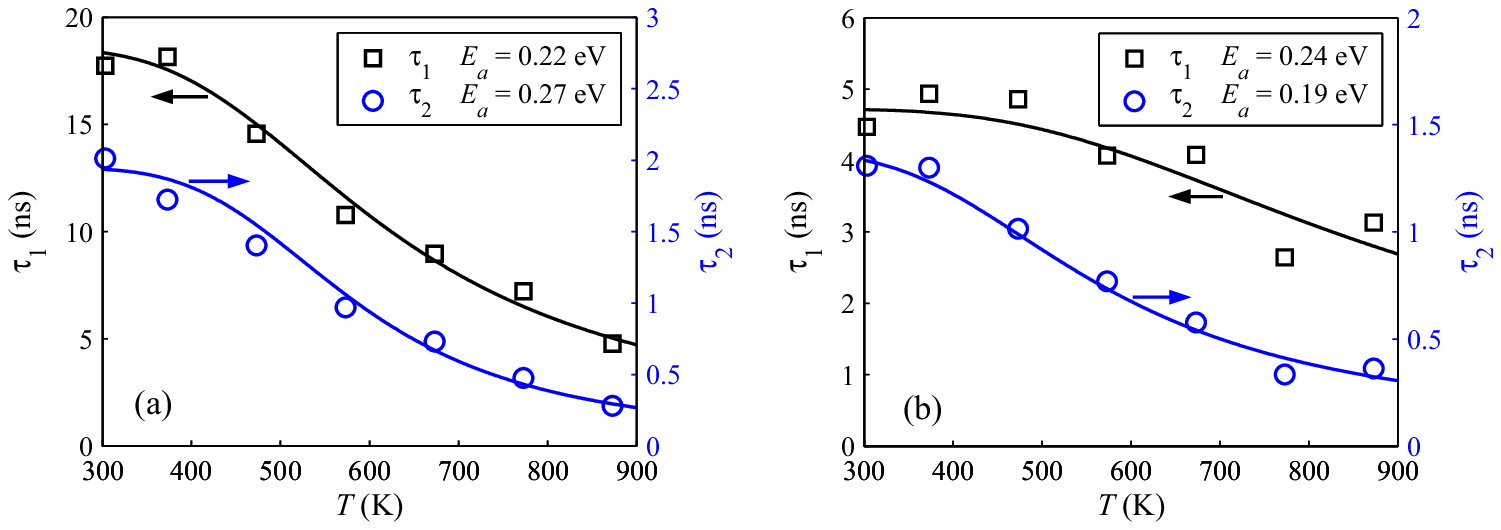}
\caption{(Color online) Characteristic PL decay times
$\tau_1=\tau_e$ (squares) and $\tau_2$ (circles), the latter at the
reference fluence of 10 mJ/cm$^2$, versus temperature $T$, for a
I-STO sample (panel a) and a N-STO sample (panel b). The lines are
best fits based on the activated behavior given by Eq.\
(\ref{eqact}). The resulting activation energies are given in the
legend.\label{figtau12}}
\end{figure}

The best-fit values of the exponential slower decay time
$\tau_1=\tau_e$ and of the faster initial decay time $\tau_2$ (at
the reference fluence of 10 mJ/cm$^2$) for a I-STO and a N-STO
samples are shown in Fig.~\ref{figtau12} (the fit procedure is
described in more detail in the next Section). As in the case of the
PL spectral amplitudes, all decay time temperature behaviors can be
fitted fairly well by the Arrhenius law given by Eq.\ (\ref{eqact}).
The resulting activation energies are given in the figure legend and
are in the range $0.2-0.3$ eV, of the same order as those obtained
from the behavior of the PL band amplitudes.

\begin{figure}[t]
\includegraphics[angle=0, scale=0.9]{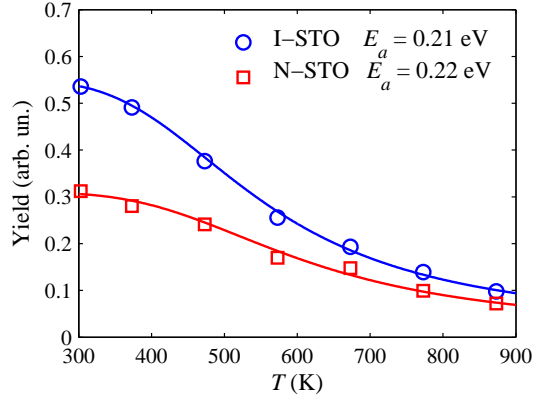}
\caption{(Color online) Photoluminescence yield, computed from the
time-integral of the decay signal, as a function of temperature, for
a I-STO and a N-STO samples. Solid lines are best fits based on the
activated behavior given by Eq.\ (\ref{eqact}). The resulting
activation energies are given in the legend.\label{figyield}}
\end{figure}

To conclude this Section, we report in Fig.~\ref{figyield} the
measured temperature behavior of the PL yields obtained from the
time integral of the decays, together with the usual activation-law
best-fits. The resulting activation energies are again roughly
consistent with the dynamical ones and with the spectral amplitude
ones. No significant dependence on doping is found in this case.

\section{Modeling the temporal decay}
Although it is firmly established that the initial decay rate of the
PL at sufficiently high excitation intensities behaves
nonlinearly,\cite{rubano07,rubano08,yasuda08} the exact decay law
and the underlying microscopic relaxation mechanisms are currently
controversial, as we mentioned in the Introduction. In this Section
we introduce and compare several different models for the PL decay,
listed in Table I, aiming at identifying the most effective one in
describing our data, which can in turn offer indications about the
underlying microscopic physics (to be discussed in the next
Section). In particular, we are especially interested in assessing
the order of the leading recombination mechanism, e.g., bimolecular
versus trimolecular.

Each tested model is labeled with a code (first column in Table I)
which synthesizes the most important assumptions on which it is
based: (i) the figure before the letter ``P'' gives the number of
decaying electron populations considered in the model (either 1 or
2), with a ``C'' denoting the more specific case of coupled
populations; (ii) the letters ``U, B, T'' stand respectively for
unimolecular, bimolecular and trimolecular recombination mechanism
(a unimolecular recombination process is taken to be present in all
models); ``v1'' and ``v2'' denote different variants of the same
basic model.

In this work, we adopted a ``global fit'' procedure, that is we
fitted simultaneously all the decays measured for a given sample at
a given temperature but for different values of the fluence $U$, for
a single choice of the adjustable parameters. For each model and for
all samples and temperatures, we have performed global best fits of
the following two kinds: (i) ``without the amplitudes'', i.e.\ on
signals which had been previously normalized to their maximum; (ii)
``with the amplitudes''. The reason for using both methods is that
the PL decay amplitudes were usually subject to a significant
scatter, not present in the decay functional form, and for the
highest excitation fluences also to some degree of saturation, which
may not be properly taken into account in our simple models.
Therefore, approach (i) is more appropriate in order to focus on the
capability of our models to predict the $U$-dependence of the PL
decay functional form, and in particular of the PL decay rates,
regardless of the amplitude behavior. Approach (ii) tests the models
in their predictive power for both PL rates and amplitudes, but
tends to weight more the amplitude behavior, as it gives the
strongest data variations. All compared models have four adjustable
parameters in approach (ii), while in approach (i) all models have
three parameters except for one (C2PUBv1), which has four. For each
approach, we quantified the performance of each model with the fit
$\chi^2$ (normalized to the data variance, estimated using the
measured noise before the excitation pulse), averaged over different
samples and repeated measurements as specified in Table I.

\begingroup
\squeezetable
\begin{table*}[t]
\centering \caption{Different PL decay models tested in this work.
The two best-fit values of $\chi^2$ reported in the last two columns
are averaged over different samples and repeated measurements, after
normalizing to the minimum $\chi^2$ value obtained among all models
for each given sample/measurement (in order to weight all samples
equally). The first $\chi^2$ (no ampl.) is computed for decays
normalized to their maximum, so that the behavior of the decay
amplitude with excitation fluence does not enter the fit and the
model testing is focused on the decay rates. The second $\chi^2$
(with ampl.) is instead computed taking also the decay amplitudes
into account. The reported differences are statistically highly
significant (the formal likelihood ratio between the best model and
the others is of thousands of orders of magnitude).}\label{table}
\begin{ruledtabular}
\begin{tabular}{l|l|l|l|l|l}
model code & rate equations & initial conditions & radiative term & $\chi^2$ no ampl.\ & $\chi^2$ with ampl.\ \\
\hline
  1PUB    & $dN/dt=-N/\tau_1-\gamma N^2$ & $N(0)=\alpha U$ & $I=Q\gamma N^2$  & 2.1 & 1.9 \\ \hline
  1PUT    & $dN/dt=-N/\tau_1-C_A N^3$ & same as above & $I=Q\gamma N^2$  & 3.2 & 1.8 \\ \hline
  2PUB    & $\begin{array}{l} dN_1/dt=-N_1/\tau_1\\ dN_2/dt=-\gamma N_2^2 \end{array}$
          & $\begin{array}{l}N_1(0)=\alpha_1U\\ N_2(0)=\alpha_2U \end{array}$ & $I=Q_1N_1/\tau_1+Q_2\gamma N_2^2$ & 2.2 & 1.7 \\ \hline
  2PUT    & $\begin{array}{l} dN_1/dt=-N_1/\tau_1\\ dN_2/dt=-C_A N_2^3 \end{array}$
          & same as above & $I=Q_1N_1/\tau_1+Q_2\gamma N_2^2$ & 3.1 & 1.7 \\ \hline
  C2PUBv1 & $\begin{array}{l} dN_1/dt=-N_1/\tau_1-\gamma_1N_1N_2\\ dN_2/dt=-\gamma_2 N_2^2 \end{array}$
          & same as above & $I=Q \left| dN_1/dt \right|$ & 1.0 & 1.7 \\ \hline
  C2PUBv2 & same as above & $\begin{array}{l} N_1(0)=\alpha_0+\alpha_1U\\ N_2(0)=\alpha_2U \end{array}$
          & $I=Q\gamma_1N_1N_2$ & 1.3 & 1.1 \\
\end{tabular}
\end{ruledtabular}
\end{table*}
\endgroup

The first two models in Table I (1PUB and 1PUT) are based on the
assumption that a single-population density $N(t)$ of decaying
electrons and holes (balanced in number) suffices for capturing the
recombination dynamics revealed by the PL decay. Many past analyses
of picosecond-laser-induced PL temporal decays in semiconductors
(see, e.g., Refs.\ \onlinecite{zarrabi85,landsberg87,linnros98}), as
well as the recent work of Yasuda \textit{et al.} on the
STO,\cite{yasuda08} interpret the PL decay by a similar single
population model. While the unimolecular and bimolecular terms are
fairly common in solid state luminescence, the appearance of a
third-order trimolecular term, typically ascribed to Auger processes
involving two electrons and a hole or two holes and an electron, is
only seen at very high excitation densities in indirect band-gap
materials. In principle, it is quite reasonable to expect this
behavior in the case of STO, and indeed the 1PUT model is that
assumed in Ref.\ \onlinecite{yasuda08}. However, we find that the
1PUT model is very ineffective in describing our data. An example of
the unsatisfactory results of a global fit based on the 1PUT model
is shown in Fig.~\ref{figconfrontodecays14STO} (gray lines, red
online), while the average best-fit $\chi^2$ values relative to the
best-performing models are reported in Table I.

We stress that by using a global (simultaneous) fit on all the
decays obtained at different excitation energies $U$, we have put
the models to the test not only on their capability to predict the
single-decay functional form, but also on their capability to
predict the overall dependence of this functional form on the
excitation fluence $U$. This dependence is very sensitive to the
order of the highest nonlinear term in the decay equations and
therefore matching this dependence provides a much stricter test
than matching the single-decay behavior. More precisely, the initial
rate of decay taking place shortly after excitation is predicted to
depend much more strongly on the excitation fluence $U$ for the 1PUT
model than for a second-order model, such as 1PUB (see Table I).
Indeed, assuming that the PL signal is generated by a given power
$p$ of the population density, $I(t)\propto N^p(t)$ (e.g., $p=2$ for
a bimolecular radiative recombination), the initial logarithmic
decay rate for a single population model including both bimolecular
and trimolecular terms is given by
\begin{eqnarray}
&& \left.-\frac{d\ln I}{dt}\right|_{t=0^+}
=\left.-\frac{p}{N}\frac{dN}{dt}\right|_{t=0^+} \nonumber\\
&& = p\left[\frac{1}{\tau_1}+\gamma N(0^+) + C_AN^2(0^+)\right]\\
&& = p\left(\frac{1}{\tau_1}+\gamma\alpha U +
C_A\alpha^2U^2\right)\nonumber
\end{eqnarray}
Thus, for the 1PUT model one would expect a quadratic dependence of
this initial decay rate on excitation fluence $U$, while for a
second-order (bimolecular) model, having $C_A\approx0$, one expects
a linear dependence. This initial decay rate is well estimated by
the inverse FW$1/e$M time $t_{1/e}^{-1}$. We indeed find a linear
behavior, as shown in Fig.~\ref{figrates}, and this supports a
bimolecular model.
\begin{figure}[t]
\includegraphics[angle=0, scale=0.9]{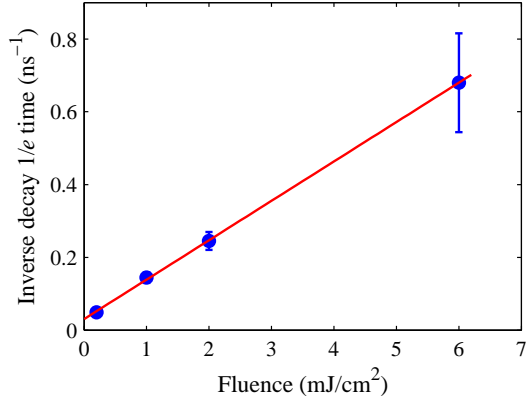}
\caption{(Color online) Inverse decay times versus excitation
fluence for the four PL decay curves reported in Fig.~1 of Ref.\
\onlinecite{yasuda08}. The times have been estimated graphically,
from the initial decay slope in the semi-logarithmic chart. The line
(red online) is a linear best-fit. The good linearity of the decay
rates vs fluence confirms a bimolecular behavior for the initial
fast decay.\label{figyasudaplot}}
\end{figure}
To investigate also the hypothesis of a crossover from a
trimolecular behavior taking place for the lower fluence range
studied by Yasuda \textit{et al.} ($0-10$ mJ/cm$^2$) to a
bimolecular one in our higher range of fluences ($10-50$ mJ/cm$^2$),
we graphically extracted from Fig.~1 of Ref.\ \onlinecite{yasuda08}
the initial logarithmic slope of the four reported decays. These
values are plotted in Fig.~\ref{figyasudaplot} versus the excitation
fluence, showing that the behavior is again perfectly linear in $U$
as expected from a bimolecular model (not necessarily a
single-population model), and hence not consistent with a
third-order model such as 1PUT. We note that the fluence dependence
of the measured decay rates can be altered by the finite response
function of the apparatus. The data from Ref.\ \onlinecite{yasuda08}
were taken with a streak camera, which typically has a very fast
response in the picosecond range, so that they should not be
affected by this problem. Our data are instead affected by the
slower response time of our equipment. Nonetheless, even after
convolution with the response function, the 1PUT third-order model
predicts a stronger dependence of the initial decay rate on the
excitation fluence than what seen in our data, as shown already in
Fig.~\ref{figconfrontodecays14STO}, and more explicitly in
Fig.~\ref{figconfronto134CRYC}.

In synthesis, we can conclude that the initial faster decay is a
second-order, or bimolecular, recombination process and no
significant trimolecular effect is detected in our data. We note
however that even at fluences much smaller than ours, indirect
semiconductors usually exhibit a strong Auger-like third-order
decay.\cite{zarrabi85,linnros98,landsberg87} Presumably, third-order
Auger interactions are depressed in STO by its relatively large
band-gap\cite{masse07} and high dielectric constant, while
bimolecular recombinations might be favored by its comparatively
large density of intragap trapping states and by its stronger
electron-phonon interactions.

\begin{figure}[t]
\includegraphics[angle=0, scale=0.9]{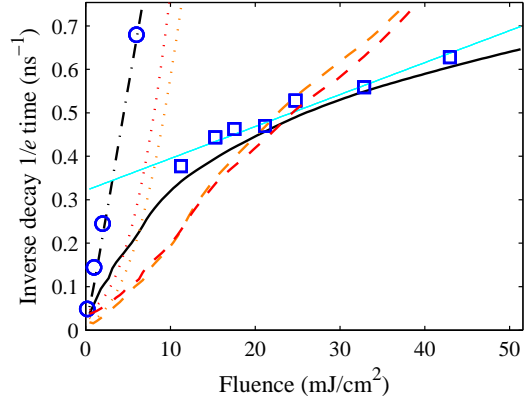}
\caption{(Color online) Inverse decay rates versus fluence as
measured for a I-STO sample (squares) and as predicted by different
models (lines), after a global best fit on the measured decays.
Besides ours, the Yasuda \textit{et al.} data already given in
Fig.~\protect\ref{figyasudaplot}) are also plotted (circles). The
two sets of data do not merge into a single smooth behavior because
of the much slower response of our apparatus as compared to that
used by Yasuda \textit{et al.} in Ref.\ \onlinecite{yasuda08}. The
predictions of the C2PUBv1 model (black lines), after a global fit
on our data only, explain both our data (solid line, obtained after
a convolution with our instrumental response function) and the
Yasuda \textit{et al.} data (dot-dash line, obtained from the
C2PUBv1 model assuming an instantaneous instrumental response).
Dashed and dotted curves correspond to the predictions of the 1PUT
model (gray line, red online) and of the 2PUT model (light gray
line, orange online), after convolution with our response function
(dashed lines) or for an instantaneous response (dotted lines). The
gray solid line (cyano online) is a linear best-fit to our data.
\label{figconfronto134CRYC}}
\end{figure}

Although it behaves much better than the trimolecular model, we see
from Table I that also the bimolecular single-population model 1PUB
is not fully satisfactory in describing our data. For this reason,
we decided to consider models based on two dynamical populations,
$N_1(t)$ and $N_2(t)$, representing for example free and trapped
charges. The simplest model of this kind is the one already
considered in our previous papers,\cite{rubano07} labeled as 2PUB,
which corresponds to the case of two independent populations, one
decaying with a unimolecular process and the other with a
bimolecular one (see Table I). The best-fits obtained using this
model are fairly good, but again not fully satisfactory. Moreover
this model does not lend itself to a simple and plausible physical
interpretation. However, this 2PUB model has the advantage of having
two well separated terms describing the initial faster decay and the
final slower tail. Therefore, it is particularly apt to describing
phenomenologically the data, for characterizing the faster and
slower decay rates of our data in a roughly model-independent way,
as we have done in Sec.\ 4. For the sake of completeness, we also
considered a two-population unimolecular + trimolecular model
(2PUT), which however can be discarded after comparison with data
(see, e.g., Fig.~\ref{figconfronto134CRYC}).

To go beyond the 2PUB model while remaining in the framework of a
two-population model, it is therefore necessary to assume some kind
of coupling between the two populations (C2PUB models). The simplest
choice is to include a term proportional to the cross product
$N_1N_2$ in one of the rate equations, i.e.\ a recombination process
of one population that is stimulated (or assisted) by the other
population. This may arise from a variety of processes, as will be
discussed in the next Section. The solution to C2PUB rate equations
(with the more general initial conditions given in the table for
C2PUBv2, see below) is the following:
\begin{eqnarray}
N_1(t)&=&\frac{(\alpha_0+\alpha_1 U)e^{-t/\tau_1}}{(1+t/\tau_2)^{\gamma_1/\gamma_2}}\nonumber\\
N_2(t)&=&\frac{\alpha_2 U}{1+t/\tau_2}. \label{eqC2PUBsol}
\end{eqnarray}
with $\tau_2=1/(\gamma_2\alpha_2U)$. The model must be now completed
with an assumption about the radiative terms. We consider here two
different possible choices for this assumption, leading to the two
model variants given in Table I. The first (C2PUBv1) is obtained
from the assumption that both recombination terms in the $N_1$
population are partly radiative, with the same quantum efficiency
$Q$. With this assumption, the predicted PL decay is given by the
following expression (in which we have also set $\alpha_0=0$, as
otherwise we would get an absurd nonzero PL for zero excitation):
\begin{eqnarray}
I(t)&=&Q\left|\frac{dN_1}{dt}\right|=k_1U\frac{e^{-t/\tau_1}}{(1+t/\tau_2)^{\gamma_1/\gamma_2}}\nonumber\\
&&+ k_2U^2 \frac{e^{-t/\tau_1}}{(1+t/\tau_2)^{1+\gamma_1/\gamma_2}}
\label{eqC2PUBfluor}
\end{eqnarray}
where $k_1$ and $k_2$ are constant amplitudes, and the following
relationship holds between the parameters:
\begin{equation}
\frac{\gamma_1}{\gamma_2}=\frac{k_2\tau_2U}{k_1\tau_1}=\frac{k_2}{k_1\tau_1\gamma_1\alpha_2}
\end{equation}
From our best fits, we obtain a typical ratio
$\gamma_1/\gamma_2=0.1-0.3$.

This model is in very good agreement with our data (for normalized
decays), as shown for example in Fig.~\ref{figconfrontodecays14STO}
and \ref{figconfronto134CRYC}. Fig.~\ref{figconfronto134CRYC} also
shows that the C2PUBv1 model, after fixing its parameters to those
giving a best fit to our data, predicts well without any further
adjustment also the decay rates measured by Yasuda \textit{et al.}
for a smaller excitation fluence range.\cite{yasuda08} The advantage
of the C2PUBv1 model, compared to the previously mentioned ones, is
also quantitatively reflected in the normalized $\chi^2$ of the fit,
which is smaller than for the other models by a factor two/three,
statistically very significant. However, when comparing normalized
decays, this model has one adjustable parameter more than the others
(four against three), so that there is still margin for doubts.
Moreover, when fitting also the PL amplitudes this model does not
perform much better than the others. This is probably revealing some
saturation behavior of the amplitude data that is not well captured
by this model.

Let us now consider the second variant of the C2PUB model (C2PUBv2)
that is obtained by the following assumption: the radiative emission
is now taken to arise only from the coupling term $N_1N_2$, i.e.\
from the decays involving both populations 1 and 2. Mathematically,
this is equivalent to setting $k_1=0$ in Eq.\ (\ref{eqC2PUBfluor}).
The number of adjustable parameters is thus back to three in the
case of normalized decays. Nevertheless, this C2PUBv2 model is
almost as effective as C2PUBv1 in the global best fits on normalized
decays. In the best-fits including also the decay amplitudes (i.e.\
in approach ii), as can be seen from the $\chi^2$ values given in
Table I, this model is actually giving by far the best results
(using the same number of adjustable parameters as for other models)
provided that we use the modified initial conditions indicated in
Table I with $\alpha_0\neq0$. This corresponds to a saturated
dependence of $N_1(0)$ versus $U$, and mathematically leads to the
replacement $k_2U\rightarrow(k_{21}U+k_{22}U^2)$ in Eq.\
(\ref{eqC2PUBfluor}) [since we know that $\gamma_1\ll\gamma_2$, we
also set $\gamma_1\approx0$ in Eq.\ (\ref{eqC2PUBfluor}), in order
to keep the number of adjustable parameters to four]. This
modification of the initial conditions does not affect the best-fits
on normalized decays because it alters only the predicted PL
amplitudes but not the decay rates and functional forms. As we will
discuss in the following Section, this modified choice of initial
conditions lends also itself to a simple and plausible physical
interpretation.

Before concluding this Section, we note that for $\tau_1\gg\tau_2$,
both models C2PUB (v1 and v2) predict an initial time decay of the
PL of the form $I(t)\sim1/(1+t/\tau_2)^{(1+\gamma_1/\gamma_2)}$,
which, for small values of the ratio $\gamma_1/\gamma_2$ such as
those found in our best fits, is very similar to that predicted by
trimolecular models, i.e.\ $I(t)\sim 1/(1+t/\tau_2)$. The other
bimolecular models 1PUB and 2PUB that we analyzed predict instead an
initial behavior that is inversely quadratic rather than linear in
the time, i.e.\ $I(t)\sim1/(1+t/\tau_2)^2$. This fact probably
explains why Yasuda \textit{et al.}, when fitting the single decay
curve, found a better fit for their decays using a trimolecular
model rather than a single population bimolecular one. On the other
hand, our global fits discriminate much more effectively among the
various possible models.

\section{Discussion and conclusions}
We now turn to discussing the microscopic physics that may underly
the STO blue photoluminescence phenomena we have described. The
photo-generated carriers may in principle populate different excited
states, which can be both localized (trapped) or extended (mobile).
Mobile charges will belong to the STO CB as electrons and to the VB
as holes, in both cases probably with some degree of phonon dressing
(large polarons). Localized charges may in principle be self-trapped
and fully intrinsic (small polarons, or self-trapped excitons if
paired) or associated with intrinsic crystalline disorder or defects
such as oxygen vacancies, dislocations, or possibly surface states.
The indirect nature of the STO bandgap forbids CB-VB direct
recombinations, so that electron coupling to phonons, presumably
enhanced by defect- or self-trapping, is an essential element of the
luminescence process.

Let us start by discussing the spectral and yield features of the
PL, in order to identify the nature of its radiative centers and of
the competing non-radiative channels that may contribute to its
quenching at high temperature. According to our fits of the PL
spectra, the typical total energy released by the annihilation of an
e-h pair is $\simeq$2.9 eV, partly dissipated in phonons and partly
radiated. The red shift of the GL spectral component with respect to
the BL and VL components is attributed to the larger fraction of
vibrational energy released in the former case, implying a stronger
electron-phonon coupling. The radiative processes giving rise to the
PL can be in principle fully intrinsic, i.e.\ characteristic of a
perfect STO crystal, such as phonon-assisted CB-VB direct
recombinations or STE annihilation, or again associated with
intrinsic structural lattice defects such as those mentioned
above.\cite{garlick67} Extrinsic defects such as chemical impurities
are likely to be excluded, instead, because the PL yield does not
present the very large sample-to-sample fluctuations that are
typical of impurity-associated luminescence (see Fig.\
\ref{fighisto}).\cite{grabner69} Moreover, the lack of any
significant doping-induced enhancement of the PL yield and of its
decay rate as seen in our intense excitation regime (see Fig.\
\ref{fighisto}), in which the doped charge density is negligible
with respect to the photoinduced one, indicates that the radiative
centers are probably not to be associated with bulk oxygen
vacancies, Nb ions or other donor centers, in contrast to what often
stated in the literature.\cite{hwang05} This is to be contrasted
with the regime of low excitation intensity, in which doping-induced
electrons are not negligible and doped samples do show a greatly
enhanced yield and decay-rate of the
PL.\cite{kan05,kan06,mochizuki05,yasuda08} The small, but
significant, sample-to-sample fluctuations of PL yield and decay
rate that we see and the presence of three different spectral
components in the PL spectra having different thermal behavior
concur to indicate that a significant role in the STO luminescence
is played by unidentified intrinsic defects (other than standard
bulk oxygen vacancies), most likely by providing the radiative
centers responsible for the luminescence itself. A less likely
alternative hypothesis is that the luminescence is a fully intrinsic
process (e.g., taking place in STEs) and the defect role is that of
introducing competing non-radiative decay channels which decrease
the yield. In both cases, the responsible defects might be titanium
interstitials,\cite{zhang08} crystal dislocations,\cite{szot06} or
other defect complexes\cite{zhang08,longo08} or, possibly, surface
defects, as suggested by the strong PL enhancement seen in
acid-etched samples and in STO nanoparticles.\cite{li07,zhang00} On
the possible role of the surface states, see also Ref.\
\onlinecite{kareev08}. However, the PL spectral difference between
Nb-doped and pure samples that we observed at high temperatures
remains unexplained.

\begin{figure}[t]
\includegraphics[angle=270, scale=0.9]{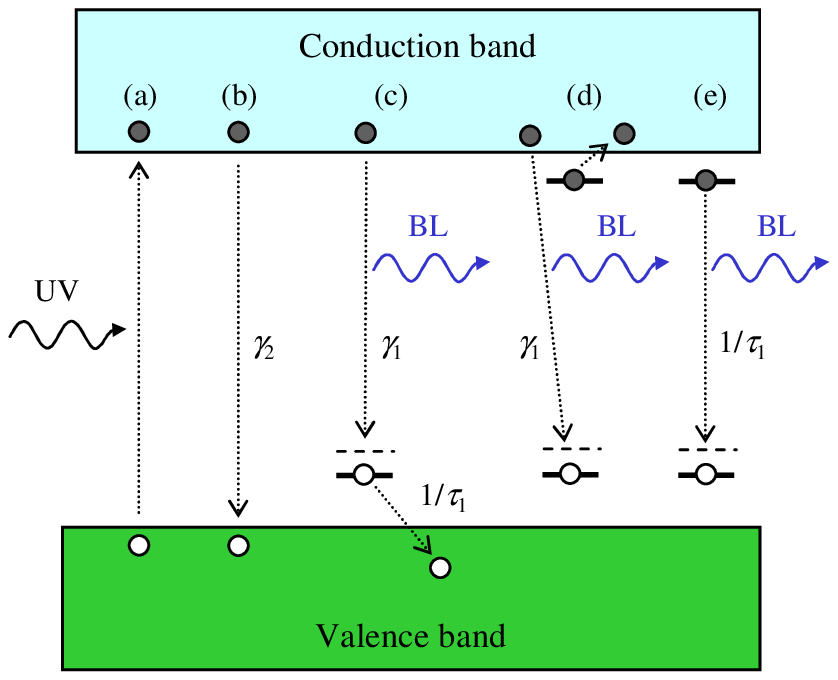}
\caption{(Color online) Schematic picture of the main electronic
processes that may be postulated for interpreting the rate equations
of our decay models C2PUB v1 and v2: (a) - pump photon (UV)
absorption and electron-hole pair generation; (b) - non-radiative
direct electron-hole recombination; (c) radiative
cross-recombination between a free electron and a trapped hole, with
blue photon (BL) emission, and non-radiative thermal untrapping of
holes; (d) - radiative crossed recombination between a free electron
and a trapped exciton; (e) - radiative spontaneous annihilation of a
trapped exciton. The constants $\gamma_1, \gamma_2, 1/\tau_1$ are
the dynamical rates of each process.\label{figmodel}}
\end{figure}

Let us now move on to discussing the physical interpretation of the
PL decay dynamics that naturally emerges from our best model, i.e.\
C2PUB (v1 or v2), with the help of Fig.~\ref{figmodel} which
provides a schematic picture of the most important dynamical
processes assumed to occur in the system. At time zero, the UV
excitation will generate many electron-hole pairs in the CB and VB
(process a), part of which will redistribute very quickly (within a
time-scale of few picoseconds, negligible for our experimental
resolution) in trapped states lying in the band-gap (horizontal
solid lines in Fig.~\ref{figmodel}). The initial conditions given in
Table I are defined by this rapid redistribution process. Most
likely the number and energy depth of these traps is not symmetrical
between holes and electrons. Here, for definiteness, we assume that
there are mainly trapped holes and free electrons, although the
converse is equally possible. Trapped holes, or possibly trapped
hole-electron pairs (i.e.\ trapped excitons) will form population 1
of model C2PUB while free electrons will correspond to population 2.
It is also assumed that $N_1 \ll N_2$, i.e., the trapped holes are
only a small fraction of the total number, so that the free
electrons and holes are approximately still balanced. These free
pairs will then recombine directly in (defect-assisted or
phonon-assisted) non-radiative bimolecular processes (process b in
Fig.~\ref{figmodel}) controlled by rate constant $\gamma_2$. This
provides the faster decay channel, entirely non-radiative and
(approximately) independent of the $N_1$ decay. The best-fit values
of the time constant $\tau_2$, combined with our estimate of the
FDCF constant, leads to an estimated band-band recombination rate
constant $\gamma_2\simeq10^{-11}$ s$^{-1}$cm$^3$. This value is
intermediate between the typical orders of magnitude found for
indirect and for direct semiconductors,\cite{rubano07} which seems a
plausible result.

Thus far the two variants of model C2PUB are essentially identical.
They are however distinguished by the interpretation attributed to
population 1 and its two decay terms. The simplest (and hence more
plausible) interpretation is that corresponding to variant C2PUBv2,
described by the two processes under the label (c) in
Fig.~\ref{figmodel}. In this case, the unimolecular term (with rate
constant $1/\tau_1$) is simply taken to be a (non-radiative) thermal
untrapping of holes (not balanced by trapping of free holes, except
for very short times after excitation, because free holes decay much
faster), while the coupling term proportional to $N_1N_2$ is taken
to be a cross recombination between a trapped hole and a free
electron, leading to the PL emission, with rate constant $\gamma_1$.
Phonons will also be created in the emission (triggered by defect
vibrational excitations, shown as dashed lines in
Fig.~\ref{figmodel}), thus explaining the PL Stokes shift and the
spectral shape discussed in Sec.\ 3. The different spectral
components (BL, GL and VL) are here ascribed to the existence of
different kinds of trapping sites. We also note that in the intense
excitation regime in which we have performed our measurements, it is
likely that the initial number of trapped holes is saturated. This
would explain quite naturally the initial conditions adopted in
C2PUBv2. On the other hand, for lower excitation energies, a non
saturated linear behavior $N_1(0){\sim}U$ should be resumed, thus
explaining also the quadratic-to-linear yield crossover reported by
Yasuda \textit{et al.}.\cite{yasuda08}

To interpret model C2PUBv1 we must assume instead that population 1
corresponds to excitons (i.e.\ e-h pairs), rather than unpaired
holes, that are either self-trapped or trapped near a structural
defect. In this case, population 1 can decay either by spontaneous
annihilation (process e in Fig.~\ref{figmodel}) or by ``crossed''
recombination of a hole of the trapped pair with a mobile electron,
leading to the emission of one PL photon and the simultaneous
freeing of the electron of the pair (process d in
Fig.~\ref{figmodel}). Both the unimolecular and the coupling terms
would then be radiative with similar quantum yield, as assumed in
model C2PUBv1. Although possible in principle, we believe that this
C2PUBv1 model scenario is somewhat less plausible than the former
(the C2PUBv2 one), both because it is more complex and because it is
difficult to justify the very rapid creation of a population of
trapped excitons as would be required by the initial conditions
assumed in the model.

In conclusion, the main results of the investigation reported in
this Article are the following: (i) we have presented strong
evidence that the radiative centers involved in the blue
luminescence cannot be associated with bulk oxygen vacancies or
other donor impurities; (ii) we have shown, nevertheless, that a
crucial role in the luminescence is played by other yet-unidentified
intrinsic structural defects, such as dislocations, defect
complexes, and possibly the surface; these defects likely provide
the actual radiative centers; (iii) we have shown that the initial
decay in the investigated excitation-intensity range is dominated by
a bimolecular process, while trimolecular processes such as Auger
are not significant; (iv) we have provided strong evidence that at
least two separate interacting photoexcited charge populations are
involved in the PL dynamics, which we interpret simply as mobile and
defect-trapped charges.


\end{document}